\documentclass[preprint,authoryear,12pt]{elsarticle}

\usepackage{epsfig}

\usepackage{amssymb}

\usepackage[ps2pdf,%
a4paper=true,%
breaklinks=true,%
colorlinks=true,%
pdfauthor={First Author et al.},%
pdftitle={Template for manuscripts in Advances in Space Research}%
]{hyperref}

\journal{Advances in Space Research}

\begin{document}

\begin{frontmatter}



\title{Exoplanetary searches with gravitational microlensing: polarization issues}


\author{Alexander F. Zakharov}
\address{Institute of Theoretical and Experimental Physics,  Moscow, 117218,
Russia} \address{ Bogoliubov Laboratory of Theoretical Physics,
JINR, Dubna, 141980 Russia} \address{ North Carolina Central
University, Durham, NC 27707,
 USA
} 

\cortext[cor]{Corresponding author.} 
\ead{zakharov@itep.ru}





\author{Gabriele Ingrosso, Francesco De Paolis, Achille A. Nucita}

\ead{ingrosso,depaolis,nucita@le.infn.it}
\author{Francesco Strafella}
\ead{strafella@le.infn.it}
\address{Dipartimento di Matematica e Fisica ``Ennio De Giorgi'',
Universit\`a del Salento, CP 193, I-73100 Lecce, Italy\\
INFN Sezione di Lecce, CP 193, I-73100 Lecce, Italy}

\author{Sebastiano Calchi Novati}
\ead{novati@sa.infn.it}

\address{
Dipartimento di Fisica ``E.R. Caianiello'', Universit\`{a} di
Salerno,
       I-84081 Baronissi (SA), Italy}
\address{       Istituto Internazionale per gli Alti Studi Scientifici (IIASS), Vietri Sul Mare (SA),
       Italy}
\author{Philippe  Jetzer} \ead{jetzer@physik.uzh.ch}
\address{Institute for Theoretical Physics,
           University of  Z\"{u}rich, Winterthurerstrasse 190,
           CH-8057,
          Z\"{u}rich, Switzerland}

\begin{abstract}
There are different methods for finding exoplanets  such as radial
spectral shifts, astrometrical measurements, transits, timing etc.
Gravitational microlensing (including pixel-lensing) is among the
most promising techniques with the potentiality of detecting
Earth-like planets at distances about a few astronomical units from
their host star or near the so-called snow line with a temperature
in the range $0-100^0$~C on a solid surface of an exoplanet. We
emphasize the importance of polarization measurements which can help
to resolve degeneracies in theoretical models. In particular, the
polarization angle could give additional information about the
relative position of the lens with respect to the source.

\end{abstract}


\end{frontmatter}

\parindent=0.5 cm


\section{Introduction}

Already before the discovery of  exoplanets \cite{Mao_91}
 showed how efficient is  gravitational microlensing as a tool to
search for extrasolar planets, including the low mass ones, even at
relatively large distances from their host stars. Later on,
observations and simulations gave the opportunity to confirm the
robustness of \cite{Mao_91}  conclusions. Exoplanets near the snow
line may be also detected with this technique as it was shown, for
instance, in Fig. 8 presented by \cite{Mao_12}.
 Moreover, in
contrast with conventional methods, such as transits and Doppler
shift measurements, gravitational microlensing gives a chance to
find exoplanets not only in the Milky Way \citep{Beaulieu_06,
Dominik_10, Zakharov_09,Zakharov_11,Zakharov_10,Wright_12, Gaudi_12,
Mao_12,Beaulieu_13}, but also in nearby galaxies, such as the
Andromeda galaxy \citep{Ingrosso_09,Ingrosso_11}, so pixel-lensing
towards M31 provides an efficient tool to search for exoplanets and
indeed an exoplanet might have been already discovered in the
PA-N2-99 event \citep{An_04, Ingrosso_09}. Since source stars for
pixel-lensing towards M31 are basically red giants (and therefore,
their typical diameters are comparable to Einstein diameters and the
caustic sizes) one has to take into account the source finiteness
effect \citep{Pejcha_09}. In the case of relatively small size
sources, the probability to have features due to binary lens (or
planet around star) in the light curves  is also small since it is
proportional to the caustic area. Giant star sources have large
angular sizes and relatively higher probability to touch caustics
\citep{Ingrosso_09}.

{

In the paper we point out an importance of polarization observations
for microlensing event candidates to support (or reject)
microlensing model and resolve degeneracies of binary (exoplanetary)
microlens models. }

\section{Exoplanet Searches with Gravitational Microlensing}

Since the existence of planets around lens stars leads to the
violation of circular symmetry of lens system and, as a result, to
the formation of fold and cusp type caustics \citep{SEF,Zakharov_95,
Petters_01}, one can detect extra peaks in the microlensing light
curve due to caustic crossing by the star source as a result of its
proper motion.

\begin{table}[t!]
\caption{Exoplanets discovered with microlensing. 24 exoplanets have
been found in 22 systems, in particular, there are two exoplanets in
OGLE-2006-BLG-109L (lines 5,6) and there are two exoplanets in
OGLE-2012-BLG-0026 (lines 18,19), see references: [1]
\cite{Bond_04},\cite{Bennett_06}; [2] \cite{Udalski_05},
\cite{Dong_09a}; [3] \cite{Beaulieu_06}; [4] \cite{Gould_06}; [5]
\cite{Gaudi_08}; [6] \cite{Gaudi_08}; [7] \cite{Sumi_10}; [8]
\cite{Bennett_08}; [9] \cite{Dong_09b}; [10] \cite{Janczak_10}; [11]
\cite{Miyake_11}; [12] \cite{Batista_11}; [13] \cite{Muraki_11};
[14] \cite{Yee_12}; [15] \cite{Bachelet_12}; [16] \cite{Bennett_12};
[17] \cite{Kains_13}; [18] \cite{Han_13a}; [19] \cite{Han_13a}; [20]
\cite{Han_13b}; [21] \cite{Tsapras_13, Poleski_13}; [22]
\cite{Furusawa_13};  [23] \cite{Choi_13}; [24] \cite{Choi_13}.}
\begin{center}
\scalebox{0.85}
 {
\begin{tabular}{|c|c|c|c|c|} \hline
\# & Star Mass  ($M_{\odot}$)  & Planet Mass & Star--planet Separation (AU) & Reference \\

\hline
& & & & \\
1&$0.63 ^{+0.07}_{-0.09}$  & $830^{+250}_{-190}M_{\oplus}$  & $4.3^{+2.5}_{-0.8}$ &[1] \\
2&$0.46 \pm {0.04}$  & $(1100 \pm 100) M_{\oplus}$  & $(3.6 \pm {0.2})$ & [2] \\
3&$0.22 ^{+0.21}_{-0.11}$  & $5.5^{+5.5}_{-2.7}M_{\oplus}$  & $2.6^{+1.5}_{-0.6}$& [3]  \\
4&$0.49 ^{+0.23}_{-0.29}$  & $13^{+6.0}_{-8.0}M_{\oplus}$  & $2.7^{+1.7}_{-1.4}$ & [4] \\
5&$0.51^{+0.05}_{-0.04}  $  & $(230 \pm 19)M_{\oplus}$  & $(2.3 \pm {0.5})$ & [5] \\
6 & $0.51^{+0.05}_{-0.04}  $  & $(86 \pm 7)M_{\oplus}$  & $4.5^{+2.1}_{-1.0}$ & [6] \\
7 & $0.64^{+0.21}_{-0.26}$  & $20^{+7}_{-8}  M_{\oplus}$  & $3.3^{+1.4}_{-0.8}$ & [7] \\
8 & $0.084^{+0.015}_{-0.012}$  & $3.2^{+5.2}_{-1.8}M_{\oplus}$  & $0.66^{+0.19}_{-0.14}$ & [8] \\
9& $0.30^{+0.19}_{-0.12}$  & $260.54^{+165.22}_{-104.85}M_{\oplus}$  & $0.72^{+0.38}_{-0.16}$/$6.5^{+3.2}_{-1.2}$ & [9] \\
10 & $0.67 \pm 0.14 $  & $28^{+58}_{-23}M_{\oplus}$  & $1.4^{+0.7}_{-0.3}$ & [10] \\
11 & $0.38^{+0.34}_{-0.18} $  & $50^{+44}_{-24}M_{\oplus}$  & $2.4^{+1.2}_{-0.6}$ & [11] \\
12 & $0.19^{+0.30}_{-0.12} $  & $2.6^{+4.2}_{-1.6}M_{\rm J}$  & $1.8^{+0.9}_{-0.7}$ & [12] \\
13 & $0.56 \pm 0.09 $  & $10.4 \pm 1.7 M_{\oplus}$  & $3.2^{+1.9}_{-0.5}$ & [13] \\
14 & $0.44^{+0.27}_{-0.17} $  & $2.4^{+1.2}_{-0.6}M_{\rm J}$  & $1.0\pm 0.1/3.5\pm 0.5$ & [14] \\
15 & $0.67^{+0.33}_{-0.13} $  & $1.5^{+0.8}_{-0.3}M_{\rm J}$  & $2^{+3}_{-1}$ & [15] \\
16 & $0.75^{+0.33}_{-0.41} $  & $3.7 \pm 2.1 M_{\rm J}$  & $8.3^{+4.5}_{-2.7}$ & [16]\\
17 & $0.26 \pm 0.11 $  & $0.53 \pm 0.21 M_{\rm J}$  & $2.72 \pm 0.75/ 1.50 \pm 0.50 $ & [17]\\
18 & $0.82 \pm 0.13  $  & $0.11 \pm 0.02 M_{\rm J}$  & $ 3.82 \pm 0.30 $  & [18] \\
19 & $0.82 \pm 0.13 $  & $0.53 \pm 0.21 M_{\rm J}$  & $4.63 \pm 0.37 $ & [19] \\
20 & $0.022 \pm 0.002 $  & $ 1.88 \pm 0.19 M_{\rm J}$  & $0.88 \pm 0.03 $ & [20] \\
21 & $0.44 \pm 0.07 $  & $ 2.73 \pm 0.43  M_{\rm J}$  & $3.45 \pm 0.26  $ & [21] \\
22 & $ 0.11 \pm 0.01 $  & $  9.2 \pm 2.2 M_{\oplus} $  & $ 0.92 \pm 0.16   $ & [22] \\
23 & $ 0.025 \pm 0.001  $  & $  9.4 \pm 0.5   M_{\rm J} $  & $ 0.19 \pm 0.01   $ & [23] \\
24 & $ 0.018 \pm 0.001  $  & $  7.2 \pm 0.5   M_{\rm J} $  & $ 0.31 \pm 0.01   $ & [24] \\
\hline
\end{tabular}
}
\end{center}
\label{Planets_ML}
\end{table}

\begin{figure}[t!]
\begin{center}
\includegraphics[angle=90,width=\textwidth]{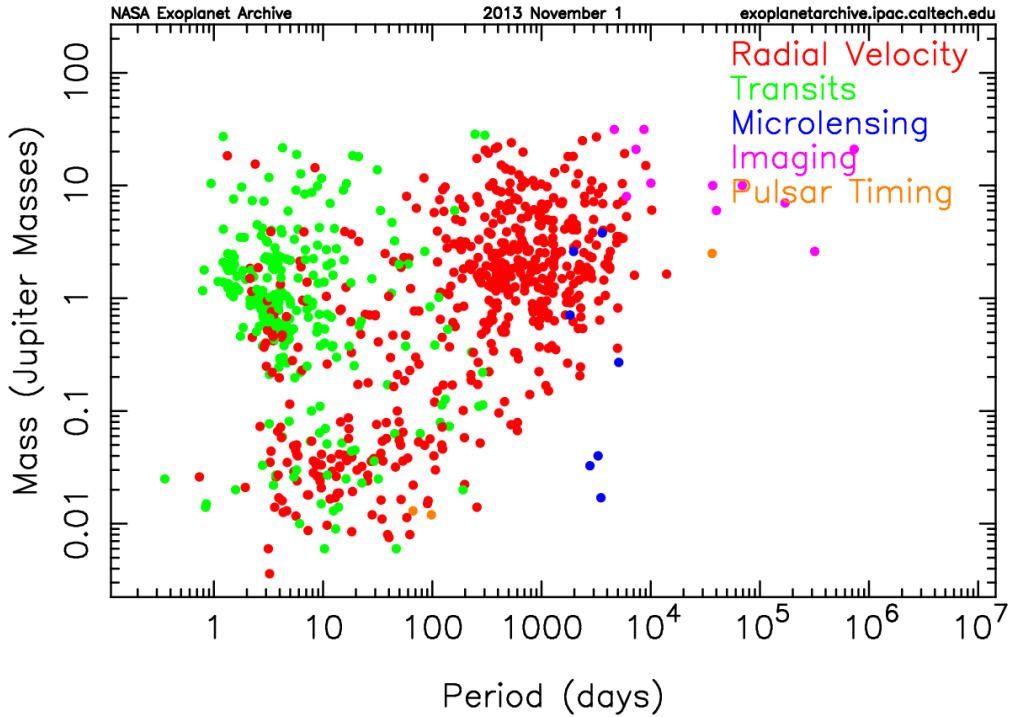}
\end{center}
\protect\caption{All exoplanets found with different techniques
until November 1, 2013, see
http://exoplanetarchive.ipac.caltech.edu/exoplanetplots/. }
\label{Exoplanets_all}
\end{figure}

A list of exoplanets detected  with microlensing searches toward the
Galactic bulge is given in Table \ref{Planets_ML} (see,
\cite{Bennett_06,Bennett_08,Bennett_09,Dong_09a,Dong_09b,Mao_12,Kains_13}).
For some planetary systems two probable regions for the
planet-to-star distance are given due to the planet and star-lens
parameter degeneracy \citep{Dominik_99,Bennett_09}, see rows 9, 14,
17 in Table \ref{Planets_ML}. Reports about these discoveries were
published  by \cite{Bond_04,Udalski_05, Beaulieu_06, Gould_06,
Gaudi_08, Bennett_08, Dong_09a,Dong_09b,
Janczak_10,Miyake_11,Batista_11, Muraki_11, Yee_12,
Bachelet_12,Bennett_12,Kains_13,Han_13a,Han_13b,Tsapras_13,
Poleski_13,Furusawa_13,Choi_13}. In these searches we have a
continuous transition from massive exoplanets to brown dwarfs, since
an analysis of the anomalous microlensing event, MOA-2010-BLG-073
has been done by \cite{Street_13}, where the primary of the lens is
an M-dwarf with $M_{L1}=0.16 \pm 0.03M_{\odot}$ while the companion
has $M_{L2}=11.0 \pm 2.0 M_{\rm J}$\footnote{According to the
definition of a "planet" done by the working group of the
International Astronomical Union on February 28, 2003 has the
following statement: "... Objects with true masses below the
limiting mass for thermonuclear fusion of deuterium (currently
calculated to be 13 Jupiter masses for objects of solar metallicity)
that orbit stars or stellar remnants are "planets" (no matter how
they formed).."}, at a perpendicular distance around $1.21 \pm 0.16$
AU from the host star, so the low mass component of the system is
near a boundary between planets and brown dwarfs.

 It is remarkable that the first exoplanet has been discovered by the MOA-I collaboration with only a 0.6~m telescope
 \citep{Bond_04,Bennett_09}. This microlensing event was also detected by the OGLE collaboration,
  but the MOA observations with a larger field of view
CCD, made about 5 exposures per night for each of their fields.
  This was an important advantage and shows that even observations with modest facilities (around 1 meter telescope size and even smaller)
  can give a crucial
contribution in such discoveries.
 Until now four super-Earth exoplanets (with masses about $10 M_{\oplus}$) have been discovered by microlensing
  (see Table \ref{Planets_ML} and Fig. \ref{Exoplanets_all}), showing that this technique  is very efficient in detecting Earth mass exoplanets
  at  a few AU from their host stars, since a significant fraction of all exoplanets discovered with different
  techniques and located in the region near the so-called snow line
  (or the habitable zone)
found with gravitational microlensing.
   Some of these exoplanets are
  among the lighest exoplanets see lines 3 and 8 in Table
  \ref{Planets_ML}.  For comparison, Doppler shift measurements help to detect
   an Earth-mass planet orbiting our neighbor star a Centauri B.
   The planet has an orbital period of 3.236 days and is about 0.04 AU
 from the star \citep{Dumusque_12}. Recently, a sub-Mercury size exoplanet Kepler-37b has been discovered
  with a transit technique \citep{Barclay_13}.
 It means that the existence of cool rocky planets is
a common phenomenon in the Universe \citep{Beaulieu_06,
Dominik_06,Dominik_06a}. Moreover, recently, \cite{Cassan_12}
claimed that around 17\% of stars host Jupiter-mass planets
($0.3-10~M_{\rm J}$), cool Neptunes ($10-30 M_{\oplus}$) and
super-Earths ($10-30 M_{\oplus}$) have relative abundances per star
in the Milky Way such as 52\% and 62\%, respectively. Analysis of
Kepler space telescope data also shows that a significant fraction
of all stars has to have exoplanets \citep{Fressin_13}.

Clearly, that if angular sizes of source stars are comparable with
corresponding angular impact parameters and  Einstein -- Chwolson
angles then light curves for such sources are different from the
standard Paczy\'nski light curve and gravitational lensing could be
colorful since one has limb darkening and color distribution for
extended background stars
\citep{Witt_94,Witt_95,Bogdanov_95a,Bogdanov_95b, Bogdanov_96a}. The
extended source effects in gravitational microlensing enable
studying the stellar atmospheres through their limb-darkening
profiles and by modelling their microlensed spectra, see
\cite{Loeb_95,Sasselov_96,Alcock_97,Sasselov_98,Heyrovsky_00,Cassan_04,Cassan_06,Thurl_06,Zub_11}
and references therein for details.


Pixel-lensing towards M31 may provide an efficient tool to search
for exoplanets in that galaxy \citep{Chung_06,Kim_07,Ingrosso_09},
and indeed an exoplanet might be already discovered in the PA-N2-99
event \citep{Ingrosso_09}. Since source stars for pixel-lensing
towards M31 are basically red giants (and therefore, their typical
diameters are comparable to Einstein diameters and the caustic
sizes) one has to take into account the source finiteness effect,
similarly to microlensing in quasars \citep{Agol_99, Popovic_06,
Jovanovic_08,Zakharov_09}. As it is well known the amplifications
for a finite source and for a point-like source are different
because there is a gradient of amplification in respect of a source
area. If the source size is rather small, the probability to produce
features of binary lens (or planet around star)  is proportional to
the caustic area. However, giant stars have  large angular sizes and
relatively higher probability to touch planetary caustics (see
\cite{Ingrosso_09}, for more details).

\section{Polarization curves for microlens systems with exoplanets}

For extended sources, the importance of polarization measurements
was pointed out by \cite{Bogdanov_96} for point-like lens and by
\cite{Agol_96} for binary lens (see also, \cite{Ignace_06}). For
point-like lens polarization could reach 0.1\% while for binary lens
it could reach a few percent since the magnification gradient is
much greater near caustics. It has been shown that polarization
measurements could resolve degeneracies in theoretical models of
microlensing events \citep{Agol_96}. Calculations of polarization
curves for microlensing events  with features in the light curves
induced by the presence of an exoplanet and observed towards the
Galactic bulge have been done \citep{Ingrosso_12,Ingrosso_13}. {We
use simple polarization and limb darkening models developed by
\cite{Chandra_50}, however, improved models are also developed
taking into account radiative transfer in spectral lines, see for
instance simulation results developed for Sun \citep{Stenflo_06}.}
  Here we emphasize that measurements of then
polarization angle could give additional information about the
gravitational microlensing model.\footnote{We call polarization
angle the angle which corresponds to a direction with the maximal
polarization.} { If polarization measurements are possible, in
principle, one could measure polarization as a function of direction
for an orientation of polarimeter and an instant for microlensing
event. If a polarimeter is fixed one has an additional function of
time to explain observational data, but if a polarimeter could be
rotated, polarization is an additional function of direction at each
instant.
 Such an information could help  to resolve degeneracies
and confirm (or disprove) microlensing models for observed
phenomena. }
 For instance, for a point-like lens the direction for the maximal
polarization at the instant when an amplification is also maximal
(which is perpendicular to the line connecting star and lens) may
allow to infer the direction of lens proper motion, thus allowing to
eventually pinpoint the lens in following observations. Even in the
case of binary lens, the orientation of polarization vector
corresponds to the orientation of the fold caustic (or more
correctly to the tangent vector to the fold caustic at the
intersection point with the path of source), provided the source
size is small enough.
\begin{figure}[t!]
\begin{center}
\includegraphics[width=\textwidth]{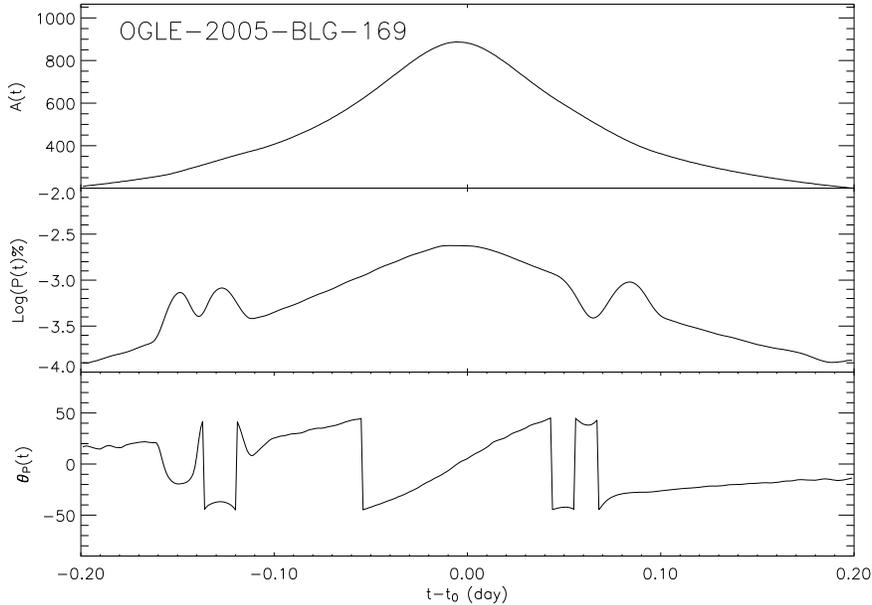}
\end{center}
\protect\caption{Light curve (top panel), polarization curve (middle
panel) and polarization angle (bottom panel) for the
OGLE-2005-BLG-169 event.} \label{Gould_06}
\end{figure}

In Fig. \ref{Gould_06},    the light curve, the polarization curve
and the polarization angle  are shown for the OGLE-2005-BLG-169
event, where a binary system formed by a main sequence star with
mass $M_\odot \sim 0.5~M_{\odot}$ and a Neptune-like exoplanet with
mass about $13~M_{\oplus}$  is expected from the light curve
analysis \citep{Gould_06}. The event parameters are
$t_E=42.27$~days, $u_0=1.24\times 10^{-3}$, $b=1.0198$, $q=8.6\times
10^{-5}$, $\alpha=117.0~$deg, $\rho_*=4.4\times 10^{-4}$, where
$t_E, u_0, b, q, \alpha, \rho_*$ are the Einstein time, the impact
parameter, the projected distance of the exoplanet to the host star,
the binary component mass ratio, the angle formed by the source
trajectory and the separation vector between the lenses, and the
source star size, respectively (all distances are given in $R_E$
units). The effect of the source transiting the caustic  (see
\cite{Gould_06} is clearly visible both in the polarization curve
(see the middle panel in Fig. \ref{Gould_06}) and in the flip of the
polarization angle (see the bottom panel). We would like to stress
that the high peak magnification ($A\simeq 800$) of the
OGLE-2005-BLG-169 event leading to $I$-magnitude of the source about
13~mag at the maximum  gives the opportunity to measure the
polarization signal for such kind of events by using present
available facilities. In this case, polarization measurements might
give additional information about the caustic structure, thus
potentially allowing to distinguish among different models of
exoplanetary systems. Recently, \cite{Gould_12} found that a
variable giant star source mimics exoplanetary signatures in the
MOA-2010-BLG-523S event. In this respect, we  emphasize that
polarization measurements may be helpful in distinguishing
exoplanetary features from other effects in the light curves.

\begin{figure}[t!]
\begin{center}
\includegraphics[width=\textwidth]{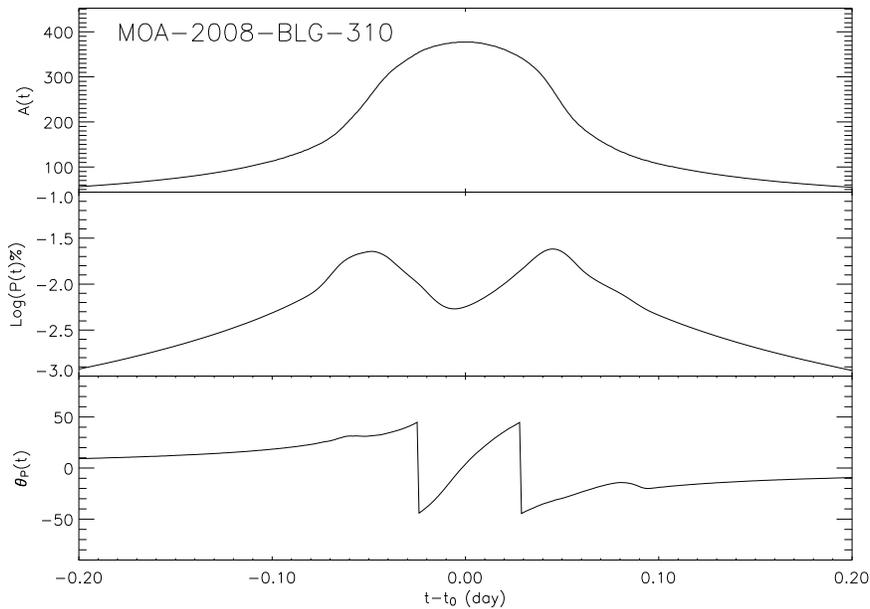}
\end{center}
\protect\caption{Light curve (top panel), polarization curve (middle
panel) and  polarization angle (bottom panel) for the
OGLE-2008-BLG-310 event.} \label{Janczak_10}
\end{figure}

The polarization curve and the polarization angle for the
MOA-2008-BLG-310Lb event  is shown in Fig. \ref{Janczak_10}. For
this event it was expected the existence of a sub-Saturn exoplanet
with  mass $m=74 \pm 17~M_{\oplus}$ \citep{Janczak_10}. The event
parameters are $t_E=11.14$~days, $u_0=3.\times 10^{-3}$, $b=1.085$,
$q=3.31\times 10^{-4}$, $\alpha=69.33~$deg, $\rho_*=4.93\times
10^{-3}$.  In particular, the event is characterized by large finite
source effect since $\rho_*/u_0> 1$, leading to polarization
features similar to those of single lens events. Nevertheless, in
this case we can see the variability in the polarization signal that
arises when the source touches the first fold caustic at $t_1 \simeq
t_0 -0.07$~days, the source enters the primary lens at $t_2 \simeq
t_0 -t_E\sqrt{\rho_*^2-u_0^2}$~days, the source exits the primary
lens at $t_3 \simeq t_0 +t_E\sqrt{\rho_*^2-u_0^2}$~days and touches
the second fold caustic $t_4 \simeq t_0 +0.09$~days (see also Fig.~4
in paper by \cite{Janczak_10}).

\section{Conclusions}

Now there are campaigns of wide field observations by Optical
Gravitational Lensing Experiment (OGLE) \citep{Udalski_03} and
Microlensing Observations in Astrophysics (MOA) \citep{Bond_01} and
a couple of follow-up observations, including
MicroFUN\footnote{http://www.astronomy.ohio-state.edu/~microfun/microfun.html.}
and PLANET\footnote{http://planet.iap.fr/.}. It is important to note
that small size (even less than one meter diameter) telescopes
acting in follow-up campaigns contributed in discoveries of light
Earth-like exoplanets and it is a nice illustration that a great
science can be done with modest facilities. As it was shown by
\cite{Ingrosso_12} polarization measurements are very perspective to
remove uncertainties in exoplanet system determination and they give
an extra proof for a conventional gravitational microlens model with
suspected exoplanets. Moreover, an orientation of polarization angle
near the maximum of polarizations (and light) curves gives
information on direction of proper motion in respect to
gravitational microlens system which
 could include exoplanet. Such an information could be important for
 possible further observations of the gravitational lens system in future.

\subsection*{Acknowledgments}

AFZ thanks organizers of IX Serbian Conference on Spectral Lines in
Astrophysics
 for their kind attention to this contribution,  the COST Action
 MP1104 "Polarization as a tool to study the Solar System and beyond" for a financial support
 and
acknowledges also a partial support of the NSF (HRD-0833184) and
NASA (NNX09AV07A) grants at NCCU (Durham, NC, USA). The authors
thank referees for a constructive criticism.





\end{document}